\documentclass[%
 reprint,
superscriptaddress,
 amsmath,amssymb,
 aps,
 prb,
]{revtex4-2}

\usepackage{graphicx} % Include figure files
\usepackage{threeparttable}
\usepackage{booktabs}
\usepackage{bm}       % bold math
\usepackage{amsthm}
\usepackage[version=4]{mhchem}  % chemical
\usepackage{siunitx}
\usepackage{color}
\usepackage{colortbl}
\usepackage{xcolor}
\usepackage{soul}
\usepackage{ulem} 
\usepackage{chngcntr}  % counter
\usepackage{titlesec}
\usepackage{hyperref} % add hypertext capabilities
\begin{document}

\preprint{APS/123-QED}

\title{CrystalFlow: A Flow-Based Generative Model for Crystalline Materials} 

\author{Xiaoshan Luo}
\affiliation{Key Laboratory of Material Simulation Methods and Software of Ministry of Education, College of Physics, Jilin University, Changchun 130012, PR~China}
\affiliation{State Key Laboratory of Superhard Materials, College of Physics, Jilin University, Changchun 130012, PR~China}

\author{Zhenyu Wang}
\affiliation{Key Laboratory of Material Simulation Methods and Software of Ministry of Education, College of Physics, Jilin University, Changchun 130012, PR~China}
\affiliation{International Center of Future Science, Jilin University, Changchun, 130012, PR~China}

\author{Qingchang Wang}
\affiliation{Key Laboratory of Material Simulation Methods and Software of Ministry of Education, College of Physics, Jilin University, Changchun 130012, PR~China}

\author{Jian Lv}
\email{lvjian@jlu.edu.cn}
\affiliation{Key Laboratory of Material Simulation Methods and Software of Ministry of Education, College of Physics, Jilin University, Changchun 130012, PR~China}

\author{Lei Wang}
\email{wanglei@iphy.ac.cn}
\affiliation{Beijing National Laboratory for Condensed Matter Physics and Institute of Physics, \\Chinese Academy of Sciences, Beijing 100190, PR~China}
%\affiliation{Songshan Lake Materials Laboratory, Dongguan, Guangdong 523808, PR~China}

\author{Yanchao Wang}
\email{wyc@calypso.cn}
\affiliation{Key Laboratory of Material Simulation Methods and Software of Ministry of Education, College of Physics, Jilin University, Changchun 130012, PR~China}

\author{Yanming Ma}
\email{mym@jlu.edu.cn}
\affiliation{Key Laboratory of Material Simulation Methods and Software of Ministry of Education, College of Physics, Jilin University, Changchun 130012, PR~China}
\affiliation{International Center of Future Science, Jilin University, Changchun, 130012, PR~China}

\date{\today}% It is always \today, today,

\begin{abstract}
Deep learning-based generative models have emerged as powerful tools for modeling complex data distributions and generating high-fidelity samples, offering a transformative approach to efficiently explore the configuration space of crystalline materials. In this work, we present CrystalFlow, a flow-based generative model specifically developed for the generation of crystalline materials. CrystalFlow constructs Continuous Normalizing Flows to model lattice parameters, atomic coordinates, and/or atom types, which are trained using Conditional Flow Matching techniques. Through an appropriate choice of data representation and the integration of a graph-based equivariant neural network, the model effectively captures the fundamental symmetries of crystalline materials, which ensures data-efficient learning and enables high-quality sampling. Our experiments demonstrate that CrystalFlow achieves state-of-the-art performance across standard generation benchmarks, and exhibits versatile conditional generation capabilities including producing structures optimized for specific external pressures or desired material properties. These features highlight the model's potential to address realistic crystal structure prediction challenges, offering a robust and efficient framework for advancing data-driven research in condensed matter physics and material science.

\end{abstract}

\maketitle

\section{\label{sec:Intro}Introduction}
The prediction of the stable arrangement of atoms within a crystal, given specific chemical compositions and external conditions, is a longstanding challenge known as crystal structure prediction (CSP)~\cite{Olson_review_2000}. This fundamental problem can be framed as a global optimization task on the potential energy surface (PES) of materials, which has far-reaching implications in the fields of physics, chemistry, and materials science, as the atomic structure of a crystal directly governs its physical and chemical properties~\cite{Wang_JCP_2014,Oganov_NRM_2019,Wang_ACR_2022}. During the past few decades, substantial advancements have been achieved in CSP, primarily attributed to the evolution of a diverse array of CSP methodologies and software, grounded in the integration of sophisticated structural sampling techniques, optimization algorithms, and quantum-mechanical calculations~\cite{Goedecker_MinimaHopping_2004,Glass_USPEX_2006,Wang_PSO_2010,Pickard_AIRSS_2011,Oakley_BasinHopping_2013,Wang_MAGUS_2023,Gusev_CombOpt_2023}. Today, these computational tools are indispensable in modern computational science, offering critical insights into the phase diagrams of condensed matter and facilitating the design of novel materials with tailored properties. This progress has led to numerous groundbreaking discoveries, such as the identification of high-pressure superhydride superconductors with record-breaking critical temperatures~\cite{Wang_CaH6_2012,Liu_PNAS_2017,Peng_PRL_2017,Drozdov_LaH10_Nature_2019}.

Despite widespread successes, CSP remains inherently challenging due to its NP-hard nature. The dimensionality of the PES increases linearly with the number of atoms in the unit cell, while the number of local minima escalates exponentially~\cite{Stillinger_PRE_1999}. These factors result in unfavorable scaling properties of CSP with respect to system size, and the pursuit of more efficient and robust methodologies for structure sampling in the vast crystal space remains a persistent endeavor within this domain.

Recent advancements in deep learning generative models have introduced transformative methodologies for understanding the underlying distributions of high-dimensional data and generating realistic samples. The remarkable capabilities of these generative models have been demonstrated in areas such as large language models~\cite{Brown_ChatGPT_2020}, image generation\cite{Rombach_StableDiffusion_2021}, and protein design~\cite{Jumper_AlphaFold2_2021}. Coupled with the rapid expansion of comprehensive materials databases, these models present a promising strategy for efficiently sampling the vast crystal space, thereby helping to address the sampling challenges in CSP. Along this line, several generative models specifically designed for crystal structures have been proposed employing variational autoencoders~\cite{Hoffmann_2019,Noh_iMatGen_2019,Court_Cond-DFC-VAE_2020,Ren_FTCP_2022,Xie_CDVAE_2022,Ye_ConCDVAE_2024,Luo_CondCDVAE_2024}, generative adversarial networks~\cite{Nouira_CrystalGAN_2018,Kim_ZeoGAN_2020,Zhao_PGCGM_2023}, diffusion models~\cite{Jiao_DiffCSP_2023,Zeni_MatterGen_2025,Yang_UniMat_2023,Jiao_DiffCSP-pp_2024,Levy_SymmCD_2024,Han_InvDesFlow_2024}, flow models~\cite{Miller_FlowMM_2024}, and autoregressive models~\cite{Xiao_SLICES_2023,Antunes_CrystalLLM_2023,Sriram_FlowLLM_2024,Cao_CrystalFormer_2024,Chen_MatterGPT_2024,Wang_SLICES-PLUS_2024,Choudhary_AtomGPT_2024,Gruver_2024,Kazeev_WyFormer_2025}. These advancements have led to improved generation performance in terms of metrics such as stability, novelty, and uniqueness of the generated samples, as well as enhanced predictive power.

Most state-of-the-art crystal generative models can be broadly categorized into two main types based on the data representation: graph-based models and string-based models~\cite{Wang_review_2024}. Graph-based models are frequently combined with diffusion/flow techniques and equivariant message passing networks. This combination is particularly effective in capturing the intrinsic symmetries of crystal structures, such as permutation, rotation and periodic translation. A prominent example in this category is the Crystal Diffusion Variational Autoencoder (CDVAE)~\cite{Xie_CDVAE_2022} which effectively integrates diffusion processes within the framework of a variational autoencoder. By utilizing SE(3) equivariant message-passing neural networks, CDVAE is able to account for key symmetries of crystals, demonstrating superior performance in terms of structural and compositional validity compared to previous models. Recent advancements along this line are represented by models such as DiffCSP~\cite{Jiao_DiffCSP_2023}, MatterGen~\cite{Zeni_MatterGen_2025}, UniMat~\cite{Yang_UniMat_2023}, and FlowMM~\cite{Miller_FlowMM_2024}, which jointly generate lattice, atomic coordinates, and/or atomic types, achieving steady improvements across various standard generative metrics. Furthermore, extensions of CDVAE have been developed for conditional generation based on specific material properties, broadening its applicability~\cite{Ye_ConCDVAE_2024,Luo_CondCDVAE_2024}. Conversely, string-based models utilized sequential tokenization of crystal structures ~\cite{Antunes_CrystalLLM_2023,Sriram_FlowLLM_2024,Choudhary_AtomGPT_2024,Gruver_2024,Xiao_SLICES_2023,Chen_MatterGPT_2024,Wang_SLICES-PLUS_2024,Cao_CrystalFormer_2024,Kazeev_WyFormer_2025}, such as standard Crystallographic Information Files (CIFs), or SLICES~\cite{Xiao_SLICES_2023}. These models typically employ transformer to learn and process the sequential data, capturing the structural information encoded in these formats. Although these models generally do not explicitly consider the symmetry of crystals, they are well-suited for scaling up to work with much larger and more diverse training data, and enabling co-training with multimodal contents, in the same fashion as large language models.

Notably, space group symmetry, a critical inductive bias for modeling crystalline materials, has been introduced into both graph-based and string-based models, such as DiffCSP++~\cite{Jiao_DiffCSP-pp_2024}, SymmCD~\cite{Levy_SymmCD_2024}, CrystalFormer~\cite{Cao_CrystalFormer_2024}, CHGFlowNet\cite{Nguyen_CHGFlowNet_2023}, and WyFormer~\cite{Kazeev_WyFormer_2025}. This advancement mitigates the challenge of generating structures with high symmetry and enables more data- and compute-efficient learning. Moreover, recent developments have realized the combination of the two types of models, leveraging the strengths of large language models in handling discrete values (atomic types) and flow models in managing continuous values (atomic positions and lattice geometry)~\cite{Sriram_FlowLLM_2024}.

In this study, we present CrystalFlow, an advanced generative model for crystalline materials that employs Continuous Normalizing Flows (CNFs)~\cite{Chen_NeuralODE_2018} within the Conditional Flow Matching (CFM) framework~\cite{Lipman_CFM_2022,Tong_OT-CFM_2023}. CrystalFlow effectively transforms a simple prior density into a complex data distribution that captures the structural and compositional intricacies of crystalline material databases. This approach simultaneously generates lattice parameters, fractional coordinates, and/or atom types for crystalline systems. By leveraging recent advancements in graph-based equivariant message-passing networks, CrystalFlow effectively incorporates the fundamental periodic-E(3) symmetries of crystalline systems, namely,  permutation, rotation, and periodic translation invariance. This symmetry-aware design ensures data-efficient learning, high-quality sampling, and flexible conditional generation, allowing the model to adapt to various constraints and requirements. Our evaluation shows that CrystalFlow achieves performance comparable to or surpassing state-of-the-art models across standard generation metrics when trained on benchmark datasets. Furthermore, when trained with appropriately labeled data, it can generate structures optimized for specific external pressures or material properties, highlighting its versatility and effectiveness in addressing realistic and application-driven CSP challenges.

\section{\label{sec:Method}Method}
\begin{figure*}[ht]
\includegraphics[width=0.80\linewidth]{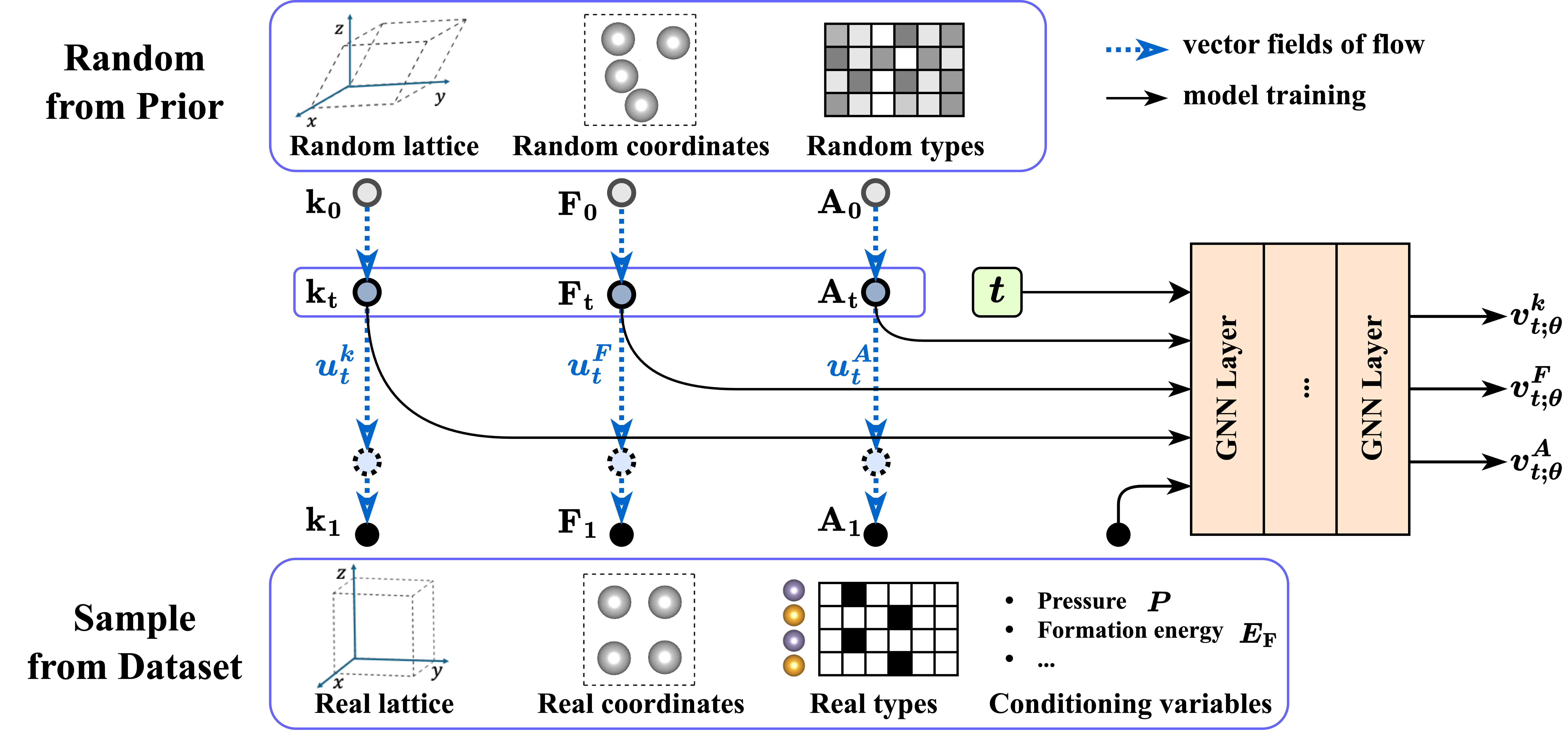}
\caption{\label{fig:fig1} \textbf{Model architecture of CrystalFlow.} Random structures, represented by lattice representations \(\mathbf{k}_0\), fractional coordinates \(\mathbf{F}_0\), and atom types \(\mathbf{A}_0\), are sampled from prior distributions. Real structures, characterized by \(\mathbf{k}_1\), \(\mathbf{F}_1\), and \(\mathbf{A}_1\), are sampled from the dataset. Continuous normalizing flows are established between these two sets, defined by vector fields \(u_t^k\), \(u_t^F\), and \(u_t^A\) at time \(t\). Intermediate structure components \(\mathbf{k}_t\), \(\mathbf{F}_t\), and \(\mathbf{A}_t\) at a given time \(t\), along with conditioning variables, serve as inputs to a graph neural network, which outputs vector fields \(v_{t;\theta}^k\), \(v_{t;\theta}^F\), and \(v_{t;\theta}^A\)\,. The model is trained by regressing the vector field \(v\) to match \(u\). For CSP tasks, \(\mathbf{A}_0 \equiv \mathbf{A}_1\) is fixed as a conditioning variable.}
\end{figure*}

In this section, we outline the foundational methodology and model architecture underlying CrystalFlow, as schematically illustrated in Fig. \ref{fig:fig1}. Following established conventions within the crystal generative modeling community, we represent a unit cell of a crystal structure containing \( N \) atoms as \(\mathcal{M} = (\mathbf{A}, \mathbf{F}, \mathbf{L})\). Here, \(\mathbf{A} = [\mathbf{a}_1, \mathbf{a}_2, \ldots, \mathbf{a}_N] \in \mathbb{R}^{a \times N}\) characterizes the chemical composition, with each atom type of a atom mapped to a unique \( a \)-dimensional categorical vector. \(\mathbf{F} = [\mathbf{f}_1, \mathbf{f}_2, \ldots, \mathbf{f}_N] \in [0,1)^{3 \times N}\) denotes the fractional coordinates of the atoms within the unit cell and \(\mathbf{L} = [\mathbf{l}_1, \mathbf{l}_2, \mathbf{l}_3] \in \mathbb{R}^{3 \times 3}\) represents the lattice matrix.

In CSP, our objective is to predict the stable structure for a given chemical composition \(\mathbf{A}\) under specific external conditions such as pressure \(P\). To achieve this, we propose designing a generative model that learns a conditional probability distribution over stable or metastable crystal configurations, denoted as \(p(x|y)\). Here \(x\) represents \( (\mathbf{F}, \mathbf{L})\), while \(y\) serves as the conditioning variable denoting \((\mathbf{A},P)\). By training this model on a comprehensive dataset of known stable and metastable structures, it can be utilized to efficiently sample high-quality crystal structures. % thereby enhancing the efficiency of CSP. 

In CrystalFlow, we model the aforementioned conditional probability distribution over crystal structures using CNF. This advanced generative modeling technique connects the data distribution \(q(x_1)\) with a simple prior distribution \(q(x_0)\), such as a Gaussian, through continuous and invertible transformations, enabling efficient sampling and exploration of complex data spaces.

\subsection{Continuous Normalizing Flow.}

The CNF framework~\cite{Chen_NeuralODE_2018} is based on a smooth, time-dependent vector field \(u:[0,1]\times\mathbb{R}^d\to\mathbb{R}^d\), which defines an ordinary differential equation (ODE):
\begin{equation}  
    \frac{dx}{dt} = u_t(x).  
\end{equation}  
\noindent The solution \(\phi_t(x)\) of this ODE, starting from \(\phi_0(x) = x\), describes the evolution of \(x\) over time. Modeling \(u_t(x)\) with a neural network \(v_{t;\theta}(t,x)\) allows us to transform a simple prior density \( p_0 \) into a complex target density \(p_1\) using the push-forward operation \(p_t=[\phi_t]_{*}p_0\) defined by:
\([\phi_t]_{*}p_0(x) := p_0(\phi_t^{-1}(x))\det\vert{\partial \phi_t^{-1}}/{\partial x}\vert\). 
The time-dependent density \(p_t(x)\) is governed by the continuity equation: \({\partial{p}}/{\partial{t}} = -\nabla\cdot({p_t}{u_t})\), ensuring conservation of probability mass. 

\subsection{Conditional Flow Matching.}

In many scenarios, the vector field \(u_t(x)\) is intractable, posing significant challenges for analysis and computation. To address this issue, CFM~\cite{Lipman_CFM_2022,Tong_OT-CFM_2023} present a simulation-free training strategy that incorporates an additional conditioning variable \(z\). This methodolegy is effective provided that a tractable conditional vector field \(u_t(x|z)\) can be defined. Suppose that the marginal probability path \(p_t(x)\) is a mixture of probatility paths \(p_t(x|z)\) that vary with some conditioning variable \(z\), that is,
\begin{equation}
p_t(x) = \int p_t(x|z)q(z)dz\;, \label{eq:marginal}
\end{equation}
where \(q(z)\) is some distribution over the conditioning variable. If \(p_t(x|z)\) evolves under a vector field \(u_t(x|z)\) from initial conditions \(p_0(x|z)\), then the vector field:
\begin{equation}
u_t(x) = \int \frac{u_t(x|z)p_t(x|z)q(z)}{p_t(x)}dz
\end{equation}
\noindent generates the probability path \(p_t(x)\) from initial conditions \(p_0(x)\). The CFM training objective becomes \(u_t(x|z)\).

This framework can be naturally extended to conditional generation with respect to a conditioning variable \(y\)~\cite{Zheng_GuidedFlow_2023}. In this case, the evolution of the system is described by: 
\begin{eqnarray}
\frac{dx}{dt} &=& u_t(x|y)\;, \label{eq:conditional.ode} \\
p_t(x|y)      &=& \int p_t(x|z)q(z|y)dz\;, \label{eq:conditional.marginal} \\
u_t(x|y)      &=&\int\frac{u_t(x|z)p_t(x|z)q(z|y)}{p_t(x)}dz\;.
\end{eqnarray}
\noindent And the training objective is still \(u_t(x|z)\).

Here, we utilize the independent coupling variant of the CFM (I-CFM)~\cite{Liu_RectifiedFlow_2022, Albergo_ICFM_2023} framework, where the conditioning variable \(z\) is defined by the pair \((x_0, x_1)\), which represent the initial and terminal points. \(p_t(x|z)\) is  a probability path interpolating between \(x_0\) and \(x_1\). The boundary conditions of Eq.~\ref{eq:conditional.marginal}  satisfy \(p_0(x|y){\approx}q(x_0|y)=q(x_0)\) being the prior distribution and \(p_1(x|y){\approx}q(x_1|y)\) being target distribution. The prior \(q(x_0)\), conditional probability path \( p_t(x|z) \) and the conditional vector field \( u_t(x|z) \) are detailed in Sec. \ref{subsec:Joint Equivariant Flow}. Finally, the training objective is written as:
\begin{equation}
\mathcal{L}_{\text{CFM}}(\theta)=\mathbb{E}_{t,q(x_1,y),q(x_0)}\Vert v_{t;\theta}(t,x,y) - u_t(x|z) \Vert^2\;,
\end{equation}
where \(v_{t;\theta}(t,x,y)\) is a time-dependent vector field parametrized as a neural network with parameters \(\theta\).

\subsection{Joint equivariant flow.}

\label{subsec:Joint Equivariant Flow}
CrystalFlow generates crystalline materials by concurrently evolving the lattice parameters \(\mathbf{L}\) and the fractional atomic coordinates \(\mathbf{F}\). In the context of crystal generative modeling, it is crucial to incorporate the fundamental periodic-E(3) symmetries of crystalline systems, namely, permutation invariance, rotation invariance, and periodic translation invariance, to enable data-efficient learning, computational efficiency, and high-quality sampling. To ensure that the push-forward distribution remains invariant under these symmetry transformations, it is generally necessary to employ an equivariant flow in conjunction with an invariant prior with respect to these symmetries~\cite{Klein_equivariant-CNFs_2023, Kohler_EquivariantFlows_2020}. In practice, achieving this invariance can be made more tractable by carefully selecting the representation of the crystal structure \(\mathcal{M}\) and designing an appropriate model architecture.

In this work, we employed a geometric graph neural network (GNN) that intrinsically accounts for permutation invariance. By utilizing a rotation-invariant representation of the lattice and the fractional coordinate system, we render the periodic-E(3) invariance tractable by fulfilling the periodic translation invariance with respect to \(\mathbf{F}\). As introduced in DiffCSP++~\cite{Jiao_DiffCSP-pp_2024}, the lattice \(\mathbf{L}\) can be represented using a rotation-invariant vector \(\mathbf{k} \in \mathbb{R}^6\), derived from the polar decomposition \(\mathbf{L} = \mathbf{Q} \exp\left(\sum_{i=1}^6 \mathbf{k}_i \mathbf{B}_i\right)\), where \(\mathbf{Q}\) is an orthogonal matrix representing rotational degrees of freedom, \(\exp(\cdot)\) is matrix exponential, and \(\mathbf{B}_i\in\mathbb{R}^{3\times 3}\) forms a standard basis over symmetric matrices. This representation effectively decouples rotational and structural information, enabling compact and symmetry-preserving parameterization of the lattice.

Let the prior distribution be \(q(x_0) = q(\mathbf{k}_0) q(\mathbf{F}_0)\),  where \( \mathbf{k}_0 \) and \( \mathbf{F}_0 \) denote the respective components of the initial state. The conditional probability path \( p_t(x | z) := p^k_t(\mathbf{k} | z) p^F_t(\mathbf{F} | z) \) is determined by the tuple \( u_t(x | z) := (u^k_t(\mathbf{k} | z), u^F_t(\mathbf{F} | z)) \). In the following, we derive the conditional probability path and the corresponding vector field for each component.

\subsubsection{Flow on lattice.} This choice of invariant lattice representation \(\mathbf{k}\) provides significant flexibility in defining both the prior distribution and the conditional probability path. For simplicity, we adopt a Gaussian prior \(q(\mathbf{k}_0) = \mathcal{N}(\mathbf{k}_0; \boldsymbol{\mu}^{k}_{0}, \boldsymbol{\sigma}^{k}_{0})\), where the mean and standard deviation are set as \(\boldsymbol{\mu}^k_{0}=(0,0,0,0,0,1)\) and \(\boldsymbol{\sigma}^k_{0} = 0.1\), respectively. The probability path is defined with a time-dependent mean \(\boldsymbol{\mu}^k_{t}\) determined by linear interpolation between the initial and terminal points, given by \(\boldsymbol{\mu}^k_{t} = t\mathbf{k}_1 + (1-t)\mathbf{k}_0\), and a constant standard deviation \(\boldsymbol{\sigma}^k_{t} = \boldsymbol{\sigma}^k\). The conditional probability path and vector field are formulated as:
\begin{eqnarray}
p^k_t(\mathbf{k}|z) &=& \mathcal{N}(\mathbf{k} | t\mathbf{k}_1 + (1-t)\mathbf{k}_0, (\boldsymbol{\sigma}^k)^2 \mathbf{I})\;, \\
u^k_t(\mathbf{k}|z) &=& \mathbf{k}_1 - \mathbf{k}_0
\label{eq:vec.field.k}\;.
\end{eqnarray}
\noindent By setting \(\boldsymbol{\sigma}^k = 0\), the conditional probability path reduces to a deterministic interpolation, consistent with the Rectified Flow framework~\cite{Liu_RectifiedFlow_2022}.

\subsubsection{Flow on fractional coordinates.} As previously discussed, the push-forward distribution \(p_t^F(\mathbf{F}|y)\), defined in Eq.~\ref{eq:conditional.marginal}, must satisfy periodic translation invariance. According to Theorem D.1 in Ref.~\cite{Miller_FlowMM_2024}, this requirement can be fulfilled by employing an invariant \(q(z|y)\) in conjunction with a pairwise invariant conditional probability path \(p_t^F(\mathbf{F}|z)\). We achieve this objective by, first, employing a uniform prior distribution \(q(\mathbf{F}_0)=\mathcal{U}(\mathbf{F}_0;0,\mathbf{I})\) guarantees \(q(z|y)\) is invariant, and second, establishing the conditional probability path as a wrapped Gaussian distribution bridging \(\mathbf{F}_0\) and \(\mathbf{F}_1\). The mean of the wrapped Gaussian, \(\boldsymbol{\mu}^F_{t}\), is determined using minimum image convention and linear interpolation, expressed as: \(\boldsymbol{\mu}^F_{t} = \mathbf{F}_0 + t (w(\mathbf{F}_1 - \mathbf{F}_0 - 0.5) - 0.5)\), where \(w(x)=x-\lfloor x \rfloor\) represents the wrapping operation that ensures periodicity. The standard deviation is taken to be constant, \(\boldsymbol{\sigma}^F_{t} = \boldsymbol{\sigma}^F\). The conditional probability path is then expressed as:
\begin{eqnarray}
&& p^F_t(\mathbf{F}|z) = \mathcal{N}_w(\mathbf{F}; \nonumber \\
&& \mathbf{F}_0 + t (w(\mathbf{F}_1 - \mathbf{F}_0 - 0.5) - 0.5), (\boldsymbol{\sigma}^F)^2 \mathbf{I})\;,
\label{eq:prob.path.f}
\end{eqnarray}
\noindent where \(\mathcal{N}_w\) represents the wrapped Gaussian distribution defined as:
\begin{equation}
\mathcal{N}_w(\mathbf{F} ; \boldsymbol{\mu}, \boldsymbol{\sigma}^2) \propto \sum_{\mathbf{Z} \in \mathbb{Z}^{3 \times N}} \exp\left[ -\frac{\Vert \mathbf{F} - \boldsymbol{\mu} + \mathbf{Z} \Vert^2}{2\boldsymbol{\sigma}^2} \right].
\label{eq:wrappedgaussian}
\end{equation}
\noindent This ensures the probability distribution to be the same over any period with interval 1, keeping the crystal periodicity.

Let the flow be the following simplest form: \(\phi^F_t(\mathbf{F})=w(\boldsymbol{\sigma}^F_{t}(z)\mathbf{F}+\boldsymbol{\mu}^F_{t}(z))\). Given that \(\boldsymbol{\sigma}^F_{t}\) is constant, and neglecting the singular points, the unique vector field is derivated as:
\begin{equation}
u^F_t(\mathbf{F}|z)=\dot{\boldsymbol{\mu}}^F_{t}(z)=w(\mathbf{F}_1-\mathbf{F}_0-0.5)-0.5\,.
\label{eq:vec.field.f}
\end{equation}

By construction, the conditional vector field \(u^F_t(\mathbf{F}|z)\) is pairwise invariant under periodic translations, resulting in the conditional probability path \(p_t^F(\mathbf{F}|z)\) being pairwise periodic translation invariant, as demonstrated in Supplementary S1. We set \(\boldsymbol{\sigma}^{F}=0\) to be consistent with the Rectified Flow framework~\cite{Liu_RectifiedFlow_2022}.

\subsubsection{Loss function.} Regressing the vector fields \(u^k_t\) in Eq.~\ref{eq:vec.field.k} and \(u^F_t\) in Eq.~\ref{eq:vec.field.f} gives rise to the total loss function \(\mathcal{L}\) used for training CrystalFlow, expressed as:
\begin{equation}
\begin{aligned}
\mathcal{L} =& \mathbb{E}_{t,q(x_1,y),q(x_0)}[\lambda_k\mathcal{L}_{k} + \lambda_F\mathcal{L}_{F}]\;, \\
\mathcal{L}_{k} =& \Vert{v_{t;\theta}^k(t,x,y) - u^k_t(\mathbf{k}|z)}\Vert^2\;, \\
\mathcal{L}_{F} =& \Vert{v_{t;\theta}^F(t,x,y) - u^F_t(\mathbf{F}|z)}\Vert^2\;,
\end{aligned}
\end{equation}
\noindent where \(\lambda_k\) and \(\lambda_F\) are weights for the respective loss terms, \(x\) represents the interpolating structure \((\mathbf{k}_t, \mathbf{F}_t)\), and \(y\) represents the chemical composition \(\mathbf{A}\) and external pressure \(P\). The vector fields \(v_{t;\theta}^k\) and \(v_{t;\theta}^F\) are parameterized by neural networks with learnable parameters \(\theta\).

\subsubsection{Flow on chemical composition.}
While CrystalFlow is primarily designed for CSP, it can be naturally extended to address inverse materials design problems, where the goal is to generate materials with specific target properties. This process is described by the generative model \( p(x|y) \), where \( x = (\mathbf{F}, \mathbf{L}, \mathbf{A}) \) and \( y \) denotes the desired properties, such as formation energy. In scenarios where \(\mathbf{A}\) (atom types) is not fixed --- a process referred to as de novo generation (DNG) --- an additional CNF is required to generate \(\mathbf{A}\) alongside the other structural components.

To generate \(\mathbf{A}\), a one-hot encoding of atomic types is employed. To reduce the dimensionality of this representation while preserving the relationships among chemical species, the periodic table is reorganized into a grid of 13 rows and 15 columns, with each species assigned a unique row-column position (detailed in Supplementary S2). The row and column indices are encoded as one-hot vectors and concatenated to represent each atomic type: \((\mathbf{a}^r,\mathbf{a}^c)\). This results in a compact 28-dimensional vector representation for each atomic type. Since \(\mathbf{A}\) is an invariant representation, the CNF applied to \(\mathbf{A}\) is analogous to that used for the lattice \(\mathbf{k}\). Specifically, we employed a Gaussian prior \(q(\mathbf{A}_0) =\mathcal{N}(\mathbf{A}_0; 0, \mathbf{I}) \in \mathbb{R}^{28}\). The conditional probability path and vector field are formulated as:
\begin{eqnarray}
p^A_t(\mathbf{A}|z) &=& \mathcal{N}(\mathbf{A} | t\mathbf{A}_1 + (1-t)\mathbf{A}_0, (\boldsymbol{\sigma}^A)^2 \mathbf{I})\;, \\
u^A_t(\mathbf{A}|z) &=& \mathbf{A}_1 - \mathbf{A}_0\;,
\end{eqnarray}
\noindent where \(\boldsymbol{\sigma}^{A}\) is also set to 0.

In comparison to the CSP task, an additional loss term for \(\mathbf{A}\) is introduced, defined as:

\begin{equation}
    \mathcal{L}_{A} = \Vert v_{t;\theta}^A(t,x,y) - u^A_t(\mathbf{A}|z) \Vert^2\;,
\end{equation}

\noindent where \(v_{t;\theta}^A\) is the predicted vector field. The total loss is given by:

\begin{equation}
\mathcal{L} = \mathbb{E}_{t,q(z,y),p_t(x|z)}[ 
\lambda_k\mathcal{L}_{k} + \lambda_F\mathcal{L}_{F} + \lambda_A\mathcal{L}_{A}]\;,
\end{equation}

\noindent where \(\lambda_k\), \(\lambda_F\), and \(\lambda_A\) are weights. The vector field \(v_{t;\theta}^A\) is parameterized by neural networks.

\subsection{Model architecture.}

A GNN is employed to predict the vector fields \(v_{t;\theta}^k\) and \(v_{t;\theta}^F\) for CSP, as well as an additional vector field \(v_{t;\theta}^A\) for DNG. Note that the periodic translational invariance with respect to \(\mathbf{F}\) should be safisfied. Specifically, we adopted the GNN architecture implemented in DiffCSP~\cite{Jiao_DiffCSP_2023}. The GNN consists of \(L\) consecutive layers, where \(\mathbf{H}^{(l)} = \{\mathbf{h}_i^{(l)} \,|\, i = 1, \ldots, N \}\) denotes the node representations at the \(l\)-th layer. The input features for node \(i\) are defined as \(\mathbf{h}_i^{(0)} = \varphi_0(f_t(t), f_A(\mathbf{a}_i)) + \varphi_y(f_y(y \setminus \mathbf{A}))\), where \(y\setminus\mathbf{A}\) represents user-specified conditions excluding the atomic composition \(\mathbf{A}\). Here, \(f_t\), \(f_A\), and \(f_y\) correspond to sinusoidal positional encoding, atomic embedding, and Gaussian basis encoding, respectively. The updates for the node feature \(\mathbf{h}_i^{(l)}\) and the edge feature \(\mathbf{m}_{ij}^{(l)}\) between nodes \(i\) and \(j\) at the \(l\)-th layer are computed as follows:
\begin{eqnarray}
\mathbf{m}_{ij}^{(l)} &=& \varphi_m(\mathbf{h}_i^{(l-1)}, \mathbf{h}_j^{(l-1)}, \mathbf{k}, \mathbf{F}^{\text{FT}}_{ij})\;, \\
\mathbf{h}_i^{(l)} &=& \mathbf{h}_i^{(l-1)} + \varphi_h\left(\mathbf{h}_i^{(l-1)}, \frac{1}{N} \sum_{j=1}^N \mathbf{m}_{ij}^{(l)}\right),
\end{eqnarray}
\noindent where \(\varphi_\Box\) represent multi-layer perceptrons (MLPs). \(\mathbf{F}^{\text{FT}}_{ij}\) denotes the Fourier transform of the relative fractional coordinate \(\mathbf{f}_j - \mathbf{f}_i\), which is periodic translation invariant. After \(L\) layers of message passing of  fully connected graph, the vector field is read out by:
\begin{eqnarray}
v_{t;\theta}^k &=& \varphi_k\left(\frac{1}{N}\sum_i^N\mathbf{h}_i^{(L)}\right) , \\
v_{t;\theta}^F &=& \varphi_F(\mathbf{h}^{(L)})\;, \\
v_{t;\theta}^A &=& \varphi_A(\mathbf{h}^{(L)})\;,
\end{eqnarray}

\noindent where \(\varphi_k\), \(\varphi_F\) and \(\varphi_A\) are also MLPs. See Supplementary S6 for other hyperparameters.

\subsection{Model inference.}

During the inference stage, a random initial state \(x_0\) --- comprising the lattice parameters, fractional atomic coordinates, and, in the case of DNG, atom types --- is sampled from the prior distribution defined in Sec.~\ref{subsec:Joint Equivariant Flow}. The final structure is subsequently determined by solving Eq.~\ref{eq:conditional.ode} over the interval \(t \in [0,1]\) using the trained vector field with a numerical ODE solver, as expressed by:

\begin{equation}
x_{1} = x_0 + \sum_{t=0}^{1} s(t)v_{t;\theta}(t, x_{t}, y)\Delta{t},
\end{equation}

\noindent where \(s(t):=1 + s^{\prime}t\) is a scaling term introduced as part of the anti-annealing numerical scheme~\cite{Miller_FlowMM_2024}, and \(\Delta{t}=1/S\), with \(S \in \mathbb{Z^+}\) representing the number of integration steps. Unless stated otherwise, we set \(S=100\), corresponding to a step size of \(\Delta{t}=1/100\). The anti-annealing parameter \(s^{\prime}\) is set to \(s^{\prime}=5\) for CSP tasks, where it is applied exclusively to \(\mathbf{F}\). For DNG tasks, the parameter is set to \(s^{\prime}=0\), effectively disabling the anti-annealing adjustment. 

\section{\label{sec:Results}Results}

In this section, we evaluate the performance of CrystalFlow on a diverse set of crystal generation tasks using datasets that span a broad range of compositional and structural diversity. The model's effectiveness is systematically benchmarked against existing crystal generation methods using standard evaluation metrics. Furthermore, the quality of the generated structures is thoroughly analyzed through detailed density functional theory (DFT) calculations.

\subsection{CSP performance on MP-20 and MPTS-52 datasets.}

\begin{table}
\caption{\label{tab:1} \textbf{CSP performance of CrystalFlow compared to previous generative models.} The match rate (MR) indicates the percentage of structures successfully reconstructed in the held-out test set, while the normalized root mean squared error (RMSE) is averaged over all matched structures. Best results are marked in bold.}
\begin{threeparttable}
\begin{tabular}{
l
S[table-format=3,detect-weight,mode=text]
S[table-format=2.2,detect-weight,mode=text]
S[table-format=1.4,detect-weight,mode=text]
S[table-format=2.2,detect-weight,mode=text]
S[table-format=1.4,detect-weight,mode=text]
}
\toprule
{method} & & \multicolumn{2}{c}{MP20}     & \multicolumn{2}{c}{MPTS52}  \\
\cmidrule(r){3-4} \cmidrule(r){5-6} & {\(k\)} & {MR (\%)} & {RMSE} & {MR (\%)} & {RMSE} \\
\midrule
CDVAE      & 1 & 33.90 & 0.1045 &  5.34 & 0.2106 \\
DiffCSP    & 1 & 51.49 & 0.0631 & 12.19 & 0.1786 \\
FlowMM     & 1 & 61.39 & \fontseries{b}\selectfont 0.0566 & 17.54 & 0.1726 \\
CrystalFlow& 1 & \fontseries{b}\selectfont 62.02 & 0.0710 & \fontseries{b}\selectfont 22.71 & 
 \fontseries{b}\selectfont 0.1548 \\
\midrule
DiffCSP    & 20 & 77.93 & \fontseries{b}\selectfont 0.0492 & 34.02 & 0.1749 \\
CrystalFlow& 20 & \fontseries{b}\selectfont 78.34 & 0.0577 & \fontseries{b}\selectfont 40.37 & \fontseries{b}\selectfont 0.1576 \\
\midrule
CrystalFlow& 100 & 82.49 & 0.0513 & 52.14 & 0.1603 \\
\bottomrule
\end{tabular}
\end{threeparttable}
\end{table}

We begin by evaluating the performance of CrystalFlow using two widely recognized benchmark datasets, MP-20 and MPTS-52~\cite{Jain_MaterialsProject_2013}. The MP-20 dataset comprises 45,231 stable or metastable crystalline materials sourced from the Materials Project (MP), encompassing the majority of experimentally reported materials in the ICSD database with up to 20 atoms per unit cell. In contrast, MPTS-52 represents a more challenging extension of MP-20, containing 40,476 crystal structures with up to 52 atoms per unit cell, organized chronologically based on their earliest reported appearance in the literature. The datasets are divided into training, validation, and test subsets in a manner consistent with previous studies. 

In accordance with standard practice, the predictive performance of the model is evaluated by calculating its match rate (MR) and  the root mean squared error (RMSE) on the test set. {Specifically, for each structure in the test set, \(k\) candidate structures are generated using CrystalFlow, and it is determined whether any of the predicted structures match the ground truth structure. The MR is defined as the fraction of structures in the test set that are successfully predicted. To match structures and evaluate their similarity, we employ the \texttt{\detokenize{StructureMatcher}} function from the Pymatgen library~\cite{Ong_pymatgen_2013}. For consistency with prior studies, the same threshold parameters are used: \texttt{\detokenize{ltol=0.3}}, \texttt{\detokenize{stol=0.5}}, \texttt{\detokenize{angle_tol=10}}. For matched structures, the RMSE between the positions of matched atom pairs is calculated and normalized by \(({\bar{V}/N})^{1/3}\), where \(\bar{V}\) is the volume derived from the average lattice parameters.}

The MR and RMSE values for CrystalFlow at \(k=1,20,\) and \(100\) on the MP-20 and MPTS-52 datasets are presented in Table~\ref{tab:1}, alongside comparisons with other crystal genevative models, including CDVAE, DiffCSP, and FlowMM. The results demonstrate that CrystalFlow achieves state-of-the-art performance, either comparable to or surpassing previous models across the evaluated metrics. On the MP-20 dataset, CrystalFlow demonstrates comparable MR and RMSE values to FlowMM, outperforming CDVAE and DiffCSP. On the more challenging MPTS-52 dataset, CrystalFlow achieves the best performance among all four models, highlighting its superior predictive capability. The time consuming for generation is given in Supplementary S7.

\subsection{CSP performance on MP-CALYPSO-60 dataset.}

\begin{figure*}
\includegraphics[width=1.0\linewidth]{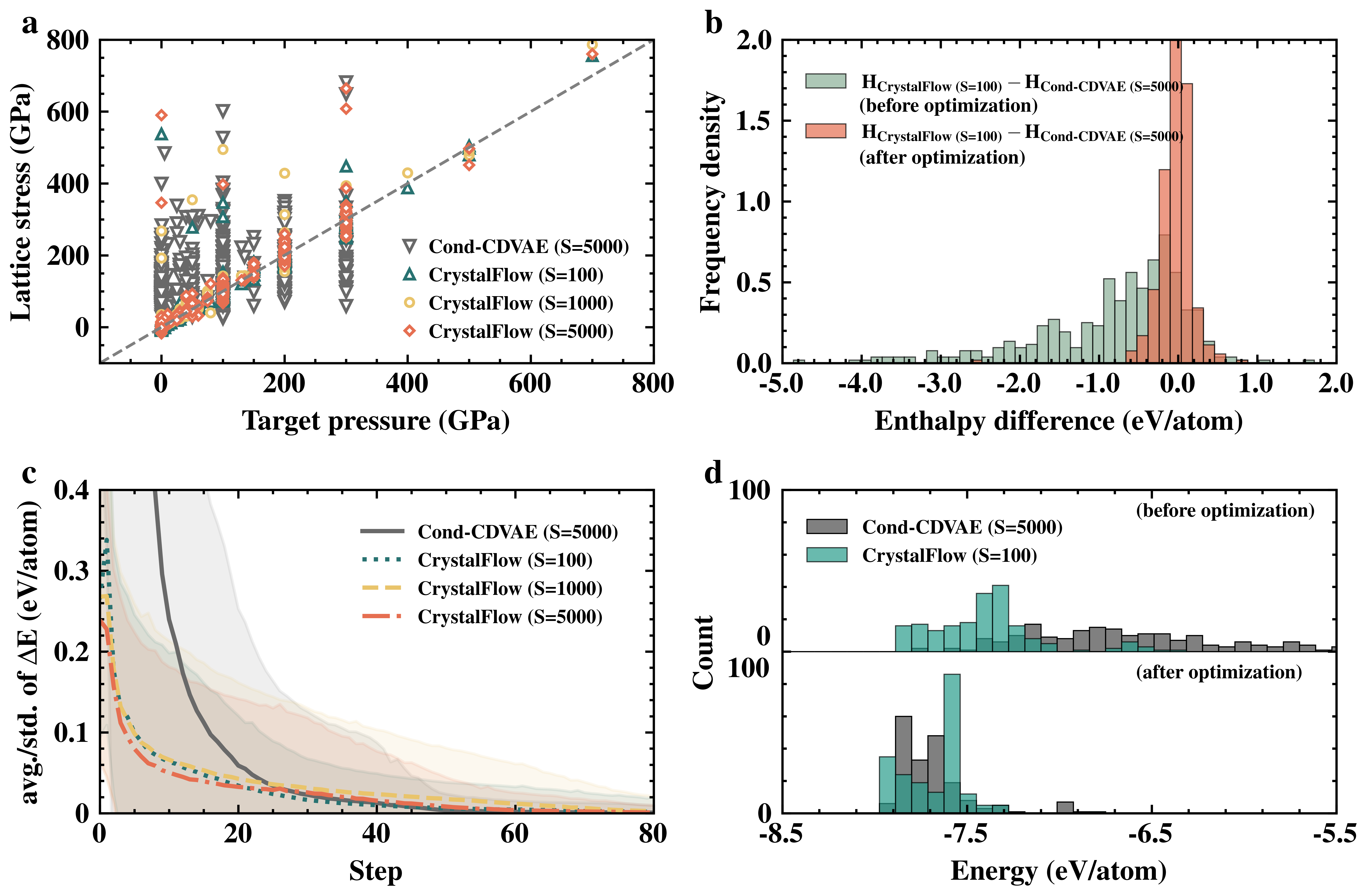}
\caption{\label{fig:2} \textbf{Performance comparison between structures generated by CrystalFlow and the previous Cond-CDVAE model.} CrystalFlow is trained on the MP-CALYPSO-60 dataset. Integration steps of \(S = 100, 1000,\) and \(5000\) are utilized for CrystalFlow, while \(S = 5000\) is employed for Cond-CDVAE. \textbf{a} The relationship between the DFT-computed lattice stress and the target pressure for 500 structures generated by each model. The composition and target pressure are randomly sampled from the test set. \textbf{b} Distributions of enthalpy differences for these structures before and after local optimization. \textbf{c} Average energy curves during local optimization for 200 \ce{SiO2} structures generated by each model at 0 GPa, with shaded areas denoting standard deviation. \textbf{d} Energy distributions of these \ce{SiO2} structures before and after local optimization.}
\end{figure*}

We further evaluate the performance of CrystalFlow using the extensive MP-CALYPSO-60 dataset described in Ref.~\cite{Luo_CondCDVAE_2024}. This dataset is constructed by integrating two sources: (1) ambient-pressure crystal structures obtained from the MP database, and (2) crystal structures generated from previous CALYPSO~\cite{Wang_CALYPSO_2012} CSP tasks conducted over a wide pressure range, with the majority of structures corresponding to pressures between 0 and 300 GPa (see Supplementary S3 for the pressure distribution). In contrast to the original dataset in Ref.~\cite{Luo_CondCDVAE_2024}, structures containing more than 60 atoms per unit cell have been excluded. The resulting dataset comprises 657,377 crystal structures, spanning 86 elements and 79,884 unique chemical compositions.  CrystalFlow, trained on this dataset, is conditioned on both chemical composition and external pressure, allowing it to generate crystal structures across a variety of pressure conditions. This capability is particularly important for simulating realistic conditions that materials may encounter in practical applications.

We generated 500 crystal structures using CrystalFlow and the previous Cond-CDVAE~\cite{Luo_CondCDVAE_2024} model for comparison, conditioned on chemical compositions and pressures randomly sampled from the test set. For CrystalFlow, integration steps \(S= 100, 1000,\) and \(5000\) were employed, whereas for Cond-CDVAE, a diffusion-based model, a large integration step of \(S=5000\) was used.  To assess the quality of the generated structures, all samples were subjected to DFT single-point calculations and local optimizations by VASP~\cite{Kresse_VASP_1996} (see Supplementary S4 for computational details). The relationship between the DFT-computed lattice stress and the target pressure specified during generation is illustrated as a scatter plot in Fig.~\ref{fig:2}a, while Fig.~\ref{fig:2}b presents the distributions of enthalpy differences between the structures generated by the two models, both before and after optimization. 

As observed in Fig.~\ref{fig:2}a, with the exception of a few outliers, the structures generated by CrystalFlow across all integration steps exhibit significantly improved alignment with the target pressures compared to those generated by Cond-CDVAE. This underscores CrystalFlow's superior ability to learn and incorporate the effects of pressure within the generative modeling process. Consequently, the majority of CrystalFlow-generated structures, even with integration steps as low as \(S=100\), exhibit lower enthalpy than those produced by Cond-CDVAE (Fig.~\ref{fig:2}b), suggesting that CrystalFlow generates more physically plausible lattice and geometric configurations. After optimization, the distribution of enthalpy differences between the two models narrows, as the optimization process mitigates the discrepancies between the initial structures. This suggests that both models are effective in learning and incorporating essential structural information from the dataset.

The convergence rate and the number of ionic steps during local optimization are essential metrics for assessing the computational efficiency of generative models in practical applications, where numerous generated structures need to be optimized. The convergence rate denotes the percentage of structures successfully optimized, while the number of ionic steps refers to the iterations needed to adjust atomic positions to minimize total energy. As shown in Table \ref{tab:2}, structures generated by CrystalFlow generally achieve a higher convergence rate compared to those generated by Cond-CDVAE. Additionally, the number of ionic steps required for CrystalFlow structures decreases with increasing integration steps, suggesting that higher integration steps improve the quality of the generated samples. At an integration step of \(S=5000\), CrystalFlow requires 39.82 average ionic steps, which is lower than the 45.91 steps needed for Cond-CDVAE, indicating a 13.3\% reduction in computational cost.

In the previous test, we generated one sample for each randomly selected chemical system from the test set. To further evaluate the model's performance on a specific system, we conducted a case study using \ce{SiO2}, a material known for its significant structural polymorphism. We generated 200 \ce{SiO2} structures at 0 GPa, each containing three formula units per unit cell (9 atoms), using both CrystalFlow with integration steps \(S = 100, 1000,\) and \(5000\), and Cond-CDVAE with \(S = 5000\). The average energy curves during DFT local optimization are presented in Fig.~\ref{fig:2}c. These curves indicate that structures generated by CrystalFlow generally exhibit lower energies compared to those generated by Cond-CDVAE, and the energy curves tend to shift downward as the integration steps increase. This demonstrates the high quality of the structures generated by CrystalFlow, which is further supported by the energy distribution of structures (with \(S=100\)) before and after optimization shown in Fig.~\ref{fig:2}d, as well as the convergence rate and number of ionic steps detailed in Table \ref{tab:2}. With an integration step of \(S=5000\), CrystalFlow requires an average of 31.99 ionic steps, which is about 27.9\% fewer than the 44.36 steps required by Cond-CDVAE.

\begin{table}
\caption{\label{tab:2} \textbf{Performance in DFT local optimizations of structures generated by CrystalFlow and Cond-CDVAE.} The convergence rate (CR) indicates the percentage of structures that were successfully optimized, while the number of ionic steps (ion-steps) refers to the iterations required to adjust atomic positions to minimize the total energy. Statistics are 500 structures for chemical systems randomly selected from the test set in MP-CALYPSO-60, and 200 structures for the \ce{SiO2} system. Best results are marked in bold.}
\begin{threeparttable}
\begin{tabular}{
l
S[table-format=4.0,detect-weight,mode=text]
S[table-format=3.2,detect-weight,mode=text]
S[table-format=2.2,detect-weight,mode=text]
S[table-format=3.2,detect-weight,mode=text]
S[table-format=2.2,detect-weight,mode=text]
}
\toprule
{method} & & \multicolumn{2}{c}{Test set} & \multicolumn{2}{c}{\ce{Si3O6}}  \\
\cmidrule(r){3-4} \cmidrule(r){5-6} & {\(S\)} & {CR (\%)} & {ion-steps} & {CR (\%)} & {ion-steps}  \\
\midrule
Cond-CDVAE    & 5000  & 82.20 & 45.91 & 96.00 & 44.36 \\
{CrystalFLow} &  100  & 89.20 & 49.84 & \fontseries{b}\selectfont 100.00 & 35.65 \\
{CrystalFlow} & 1000  & 90.20 & \fontseries{b}\selectfont 39.40 & \fontseries{b}\selectfont 100.00 & 35.84 \\
{CrystalFLow} & 5000  & \fontseries{b}\selectfont 90.60 & 39.82 & \fontseries{b}\selectfont 100.00 & \fontseries{b}\selectfont 31.99 \\
\bottomrule
\end{tabular}
\end{threeparttable}
\end{table}

\subsection{DNG performance on MP-20.}

\begin{table*}
\caption{\label{tab:3} \textbf{Performance comparison of CrystalFlow in DNG taks with previous generative models.} Statistic were performed on 10,000 randomly generated structures. The performance are evaluated in terms of structural validity, compositional validity, coverage recall, coverage precision, and property statistics, including density (\(\rho\)) and the number of elements (\(N_\text{el}\)). Best results are marked in bold.}
\begin{threeparttable}
\begin{tabular}{
l
S[table-format=3.2,detect-weight,mode=text]
S[table-format=2.2,detect-weight,mode=text]
S[table-format=2.2,detect-weight,mode=text]
S[table-format=2.2,detect-weight,mode=text]
S[table-format=1.3,detect-weight,mode=text]
S[table-format=1.3,detect-weight,mode=text]
}
\toprule
& \multicolumn{2}{c}{Validity (\%) \(\uparrow\)}  & \multicolumn{2}{c}{Coverage (\%) \(\uparrow\)}  & \multicolumn{2}{c}{Property \(\downarrow\)} \\
\cmidrule(r){2-3} \cmidrule(r){4-5} \cmidrule(r){6-7} & {Structural} & {Compositional} & {Recall} & {Precision} & {wdist (\({\rho}\))} & {wdist (\({N_{\text{el}}}\))} \\
\midrule
CDVAE       & \fontseries{b}\selectfont 100.00 & 86.70 & 99.15 & 99.49 & 0.688 & 1.432 \\
DiffCSP     & \fontseries{b}\selectfont 100.00 & 83.25 & \fontseries{b}\selectfont 99.71 & 99.76 & 0.350 & 0.340 \\
FlowMM      &  99.85 & 83.19 & 99.49 & 99.58 & 0.239 & \fontseries{b}\selectfont 0.083 \\
FlowLLM     &  99.94 & \fontseries{b}\selectfont 90.84 & 96.95 & 99.82 & 1.140 & 0.150 \\
CrystalFlow &  99.55 & 81.96 & 98.21 & \fontseries{b}\selectfont 99.84 & \fontseries{b}\selectfont 0.169 & 0.259 \\
\bottomrule
\end{tabular}
\end{threeparttable}
\end{table*}

\begin{figure}
\includegraphics[width=1.0\linewidth]{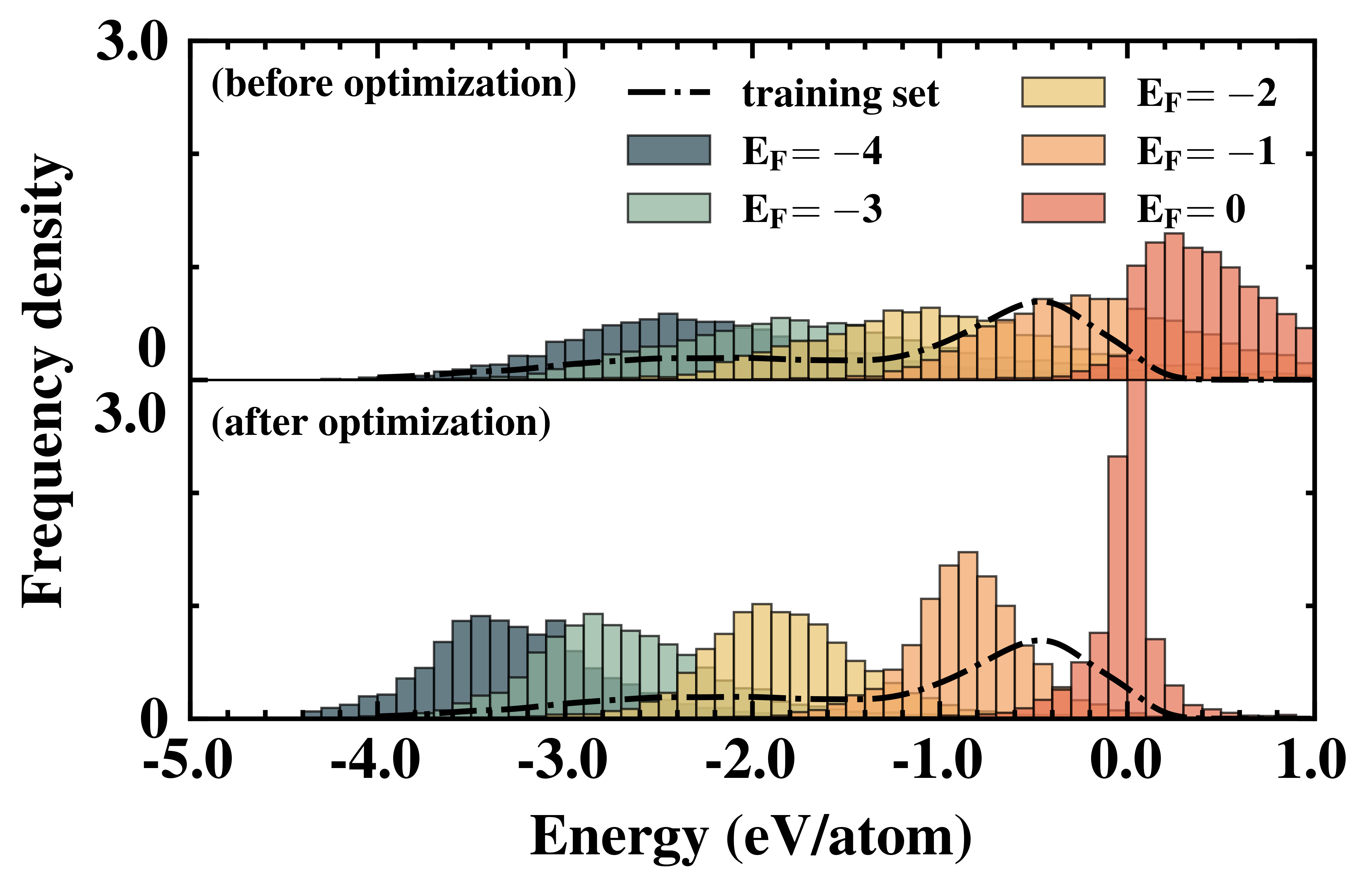}
\caption{\label{fig:3}\textbf{Performance of CrystalFlow in DNG tasks with targeted properties.} Distributions of formation energy (\(E_\text{F}\)) for CrystalFlow-generated structures, conditioned on target values of \(E_\text{F} = 0, -1, -2, -3,\) and \(-4\) eV/atom. The distributions are shown for structures (top) before and (bottom) after geometric optimization. For each target, 10,000 structures are generated. The dotted curve denotes the corresponding distribution of the training set.}
\end{figure}

Finally, we evaluate the DNG performance of CrystalFlow on the MP-20 dataset to assess its potential for inverse materials design tasks. Initially, we train CrystalFlow on MP-20 without conditioning, and compare its performance with other models using common DNG metrics, including structural and compositional validity, coverage, and property statistics. The structural validity is defined as the percentage of generated structures in which all pairwise atomic distances exceed 0.5~\AA. The compositional validity, on the other hand, refers to the percentage of generated structures with an overall neutral charge, calculated using SMACT~\cite{Davies_SMACT_2019}. Coverage quantifies the structural and compositional similarity between the test set and the generated structures with a detailed definition given in Supplementary S5. For property statistics, we evaluate the similarity between the test set and the generated structures in terms of density \(\rho\) and number of elements \(N_\text{el}\), using Wasserstein distances (wdist). 

The statistic results of 10,000 randomly generated structures are presented in Table \ref{tab:3}, and demonstrate that CrystalFlow achieves a performance comparable to that of existing models across various metrics. Although its compositional validity is marginally lower than that of other approaches, CrystalFlow achieves the smallest wdist for density, underscoring its capability to generate structures with more physically reasonable lattice parameters. 

To demonstrate CrystalFlow's capability to generate structures with targeted properties, the model is further trained on the MP-20 dataset with formation energy (\(E_{\text{F}}\)) as a conditioning label. Subsequently, 10,000 structures are generated for each specified formation energy value, conditioned on \(E_{\text{F}} = 0, -1, -2, -3,\) and \(-4\) eV/atom, respectively. To facilitate efficient evaluation, energy calculations and geometric optimization are performed using CHGNet~\cite{Deng_CHGNet_2023}, a universal and computationally efficient interatomic potential. The distributions of \(E_\text{F}\) for the generated structures, both before and after geometric optimization, are shown in Fig.~\ref{fig:3} top and bottom, respectively. The generated structures generally exhibit an \(E_\text{F}\) distribution centered around the specified target values, although with a noticeable shift toward higher-energy regions. After geometric optimization, the distributions become more closely aligned with the target values. Notably, for target \(E_\text{F}\) values with greater representation in the training dataset, the generated distributions exhibit closer agreement with the target values. These results highlight the effectiveness of CrystalFlow in generating structures with targeted properties and its ability to leverage training data to improve accuracy of conditional generation.

\section{Discussion}
By integrating state-of-the-art generative modeling techniques with domain-specific knowledge in physical and material science, we introduce CrystalFlow, a crystal generative model specifically designed for the efficient exploration of the vast and complex materials space. CrystalFlow exhibits strong capabilities in generating high-quality crystal structures that adhere to fundamental physical principles and satisfy target constraints, as demonstrated by its superior performance across multiple benchmark datasets. These attributes position CrystalFlow as a highly promising and versatile structure-sampling tool, with significant potential for integration into a wide range of CSP methods and materials design workflows.

Flow-based models offer several advantages over diffusion-based approaches, including fewer integration steps and more flexible choices of prior distributions, making them particularly appealing for generative modeling in materials science. However, as this work represents an initial step in exploring these capabilities, the full potential of flow-based models remains to be comprehensively investigated. Future studies could focus on evaluating the influence of integration strategies, integration lengths, and prior distribution choices on the quality and diversity of generated structures. Moreover, incorporating fixed structural modifications or enforcing space group symmetries during generation could further enhance the model's ability to produce realistic and physically meaningful crystal structures.

Assessing the performance of CrystalFlow on larger and more diverse datasets, as well as under multi-property constraints, would provide valuable insights into its scalability, robustness, and applicability to real-world materials discovery. Furthermore, integrating CrystalFlow into broader generative frameworks --- such as hybrid architectures that combine flow-based models with autoregressive approaches --- holds significant promise for further improving its generative performance. These directions represent exciting opportunities to advance the capabilities of crystal generative modeling.

We anticipate that this work will inspire further innovations at the intersection of artificial intelligence, condensed matter physics, and material science, ultimately advancing the discovery and design of next-generation materials.

\section{Data availability}

The authors declare that the main data supporting the findings of this study are contained within the paper and its associated Supplementary Information.

\section{Code availability}

The CrystalFlow source code is available on GitHub (\href{https://github.com/ixsluo/CrystalFlow}{https://github.com/ixsluo/CrystalFlow}).

% \nocite{*}

\bibliographystyle{naturemag}  % apsrev4-2 naturemag
\bibliography{ref}% Produces the bibliography via BibTeX.

\begin{acknowledgments}
The work is supported by the National Key Research and Development Program of China (Grant No. 2022YFA1402304), the National Natural Science Foundation of China (Grants No. 12034009, 12374005, 52288102, 52090024, and T2225013), the Fundamental Research Funds for the Central Universities, the Program for JLU Science and Technology Innovative Research Team. We want to thank ``Changchun Computing Center" and ``Eco-Innovation Center" for providing inclusive computing power and technical support of MindSpore during the completion of this paper and the support provided by MindSpore Community. Part of the calculation was performed in the high-performance computing center of Jilin University. We also want to thank all CALYPSO community contributors for providing the structural data to construct the dataset.
\end{acknowledgments}

\section*{Author contributions}
J.L., L.W., Y.W., and Y.M. designed the research. X.L. and Q.W. wrote the code and trained the model. Z.W. and X.L. collected the dataset. X.L., Z.W. and Q.W. conducted the calculations. X.L., Z.W., Q.W., J.L., L.W., Y.W., and Y.M. analyzed and interpreted the data, and contributed to the writing of the paper.

\section*{Competing interests}

The authors declare no competing interests.

\section*{Additional information}

\textbf{Supplementary information} The online version contains supplementary material available at XXX.

\textbf{Correspondence} and requests for materials should be addressed to Jian Lv, Lei Wang, Yanchao Wang, or Yanming Ma.

%------------- for concat SI ---------------
\clearpage
\begin{widetext}
\setcounter{section}{0}
\renewcommand{\thesection}{S\arabic{section}}
\titleformat{\section}[hang]{\normalfont\Large\bfseries}{\thesection}{1em}{}
\counterwithin{figure}{section} % 图表计数器与章节关联
\counterwithin{table}{section}  % 表格计数器与章节关联
%\counterwithin{equation}{section}
\renewcommand*{\figurename}{Supplementary Fig.}   % <-- Figure style
\renewcommand*{\tablename}{Supplementary Table}   % <-- Figure style
\renewcommand*{\thefigure}{S\arabic{section}--\arabic{figure}}      % <-- Figure style
\renewcommand*{\thetable}{S\arabic{section}--\arabic{table}}      % <-- Figure style
\renewcommand*{\theequation}{S\arabic{equation}}

\begin{center}
\huge{\textbf{Supplementary Information}}
\end{center}

\section{\label{sec:s1}\texorpdfstring{Symmetry of conditional vector field \(u_t^F\) and conditional probability path \(p_t^F\) for fractional coordinates}{}}

Below, we demonstrate that the conditional velocity field \(u_t^F\), as defined in Eq.~12 of the main text, is pairwise invariant under periodic translations. This invariance ensures that the conditional probability path \(p_t^F(\mathbf{F}|z)\), as defined in Eq.~10, is also pairwise invariant under periodic translations. 

Specifically, let \(\boldsymbol{\tau} := \hat{\boldsymbol{\tau}}\mathbf{1}^{\top}\), where \(\hat{\boldsymbol{\tau}} \in \mathbb{R}^{3 \times 1}\) is an arbitrary translation vector and \(\mathbf{1} \in \mathbb{R}^{N \times 1}\) is a column vector of ones. The periodic translation operation applied to the fractional coordinates \(\mathbf{F}\) is then expressed as \(g \circ \mathbf{F} := w(\mathbf{F} + \boldsymbol{\tau})\). We now demonstrate the pairwise periodic translation invariance of \(u_t^F\) under this operation as follows:

\begin{eqnarray}
&&u^F_t(g \circ \mathbf{F} | g \circ z) \nonumber \\
&=& w(w(\mathbf{F}_1 + \boldsymbol{\tau}) - w(\mathbf{F}_0 + \boldsymbol{\tau}) - 0.5) - 0.5 \nonumber \\
&=& w(\mathbf{F}_1 + \boldsymbol{\tau} - \lfloor\mathbf{F}_1 + \boldsymbol{\tau}\rfloor - \mathbf{F}_0 - \boldsymbol{\tau} + \lfloor\mathbf{F}_0 + \boldsymbol{\tau}\rfloor - 0.5) - 0.5 \nonumber \\
&=& w(\mathbf{F}_1 - \mathbf{F}_0 - 0.5) - 0.5 \nonumber \\
&=& u^F_t(\mathbf{F}|z)\;.
\end{eqnarray}

Next, we demonstrate the pairwise periodic translation invariance of the conditional probability path \(p_t^F(\mathbf{F}|z)\), as defined in Eq.~10. Reformulating \(p_t^F(\mathbf{F}|z)\) as:
\[
p^F_t(\mathbf{F}|z) = \mathcal{N}_w(\mathbf{F} | \mathbf{F}_0 + t u^F_t(\mathbf{F}|z), \boldsymbol{\sigma}_{\mathbf{F}}^2 \mathbf{I})\;,
\]
where \(\mathcal{N}_w\) denotes the wrapped normal distribution, we proceed with the proof as follows:

\begin{eqnarray}
&& p^F_t(g \circ \mathbf{F} | g \circ z) \nonumber \\
&=& \mathcal{N}_w(g \circ \mathbf{F} ; g \circ \mathbf{F}_0 + t u^F_t(g \circ \mathbf{F} | g \circ z), \boldsymbol{\sigma}_{\mathbf{F}}^2 \mathbf{I}) \nonumber \\
&=& \mathcal{N}_w(g \circ \mathbf{F} ; w(\mathbf{F}_0 + \boldsymbol{\tau}) + t u^F_t(\mathbf{F}|z), \boldsymbol{\sigma}_{\mathbf{F}}^2 \mathbf{I}) \nonumber \\
&=& \mathcal{N}_w(w(\mathbf{F} + \boldsymbol{\tau}) ; w(\mathbf{F}_0 + \boldsymbol{\tau} + t u^F_t(\mathbf{F}|z)), \boldsymbol{\sigma}_{\mathbf{F}}^2 \mathbf{I}) \nonumber \\
&=& \mathcal{N}_w(\mathbf{F} + \boldsymbol{\tau} ; \mathbf{F}_0 + \boldsymbol{\tau} + t u^F_t(\mathbf{F}|z), \boldsymbol{\sigma}_{\mathbf{F}}^2 \mathbf{I}) \nonumber \\
&=& \mathcal{N}_w(\mathbf{F} ; \mathbf{F}_0 + t u^F_t(\mathbf{F}|z), \boldsymbol{\sigma}_{\mathbf{F}}^2 \mathbf{I}) \nonumber \\
&=& p^F_t(\mathbf{F}|z)\;.
\end{eqnarray}

Thus, the conditional probability path \(p_t^F(\mathbf{F}|z)\) is shown to be invariant under pairwise periodic translations. This result relies on the properties of the wrapped normal distribution, as established in Lemma 3 of Jiao et al.~\cite{Jiao_DiffCSP_2023}.

\clearpage
\section{\label{sec:s2}Encoding of atom type for de novo generation}

In de novo generation tasks, generating the chemical composition \(\mathbf{A}\) is a critical step. We employ a one-hot encoding scheme for atomic types to achieve this. To reduce the dimensionality of this representation while maintaining the inherent relationships among chemical species, the periodic table is reorganized into a grid consisting of 13 rows and 15 columns. Each chemical species is assigned a unique position within this grid, defined by specific row and column indices, as illustrated in Table \ref{tab:s2-1}. These indices are then encoded as one-hot vectors, which are concatenated to form a composite representation for each atomic type: \((\mathbf{a}^r, \mathbf{a}^c)\). This approach results in a compact 28-dimensional vector representation for each atomic type, effectively capturing essential chemical relationships in a reduced-dimensional space.

\begin{table}[!h]
    \caption{\label{tab:s2-1} Reorganized periodic table. The elements of sub family, lanthanides and actinides are positioned in the bottom.}
    \begin{threeparttable}
    \begin{tabular}{r|c*{14}c}
        \toprule
        & 0 & 1 & 2 & 3 & 4 & 5 & 6 & 7 & 8 & 9 & 10 & 11 & 12 & 13 & 14 \\ 
        \midrule
        0 &  H & -- & -- & -- & -- & -- & -- & He & -- & -- & -- & -- & -- & -- & -- \\ 
        1 & Li & Be &  B &  C &  N &  O &  F & Ne & -- & -- & -- & -- & -- & -- & -- \\ 
        2 & Na & Mg & Al & Si &  P &  S & Cl & Ar & -- & -- & -- & -- & -- & -- & -- \\ 
        3 & K  & Ca & Ga & Ge & As & Se & Br & Kr & -- & -- & -- & -- & -- & -- & -- \\ 
        4 & Rb & Sr & In & Sn & Sb & Te &  I & Xe & -- & -- & -- & -- & -- & -- & -- \\ 
        5 & Cs & Ba & Tl & Pb & Bi & Po & At & Rn & -- & -- & -- & -- & -- & -- & -- \\ 
        6 & Fr & Ra & Nh & Fl & Mc & Lv & Ts & Og & -- & -- & -- & -- & -- & -- & -- \\ 
        \midrule
        7 & Sc & Ti &  V & Cr & Mn & Fe & Co & Ni & Cu & Zn & -- & -- & -- & -- & -- \\ 
        8 &  Y & Zr & Nb & Mo & Tc & Ru & Rh & Pd & Ag & Cd & -- & -- & -- & -- & -- \\ 
        9 & -- & Hf & Ta &  W & Re & Os & Ir & Pt & Au & Hg & -- & -- & -- & -- & -- \\ 
       10 & -- & Rf & Db & Sg & Bh & Hs & Mt & Ds & Rg & Cn & -- & -- & -- & -- & -- \\ 
        \midrule
        11 & La & Ce & Pr & Nd & Pm & Sm & Eu & Gd & Tb & Dy & Ho & Er & Tm & Yb & Lu \\ 
        12 & Ac & Th & Pa & U & Np & Pu & Am & Cm & Bk & Cf & Es & Fm & Md & No & Lr \\ 
        \bottomrule
    \end{tabular}
    \end{threeparttable}
\end{table}

\clearpage
\section{\label{sec:s3}Pressure distribution for MP-CALYPSO-60 dataset}

\begin{figure}[!h]
\includegraphics[width=0.9\linewidth]{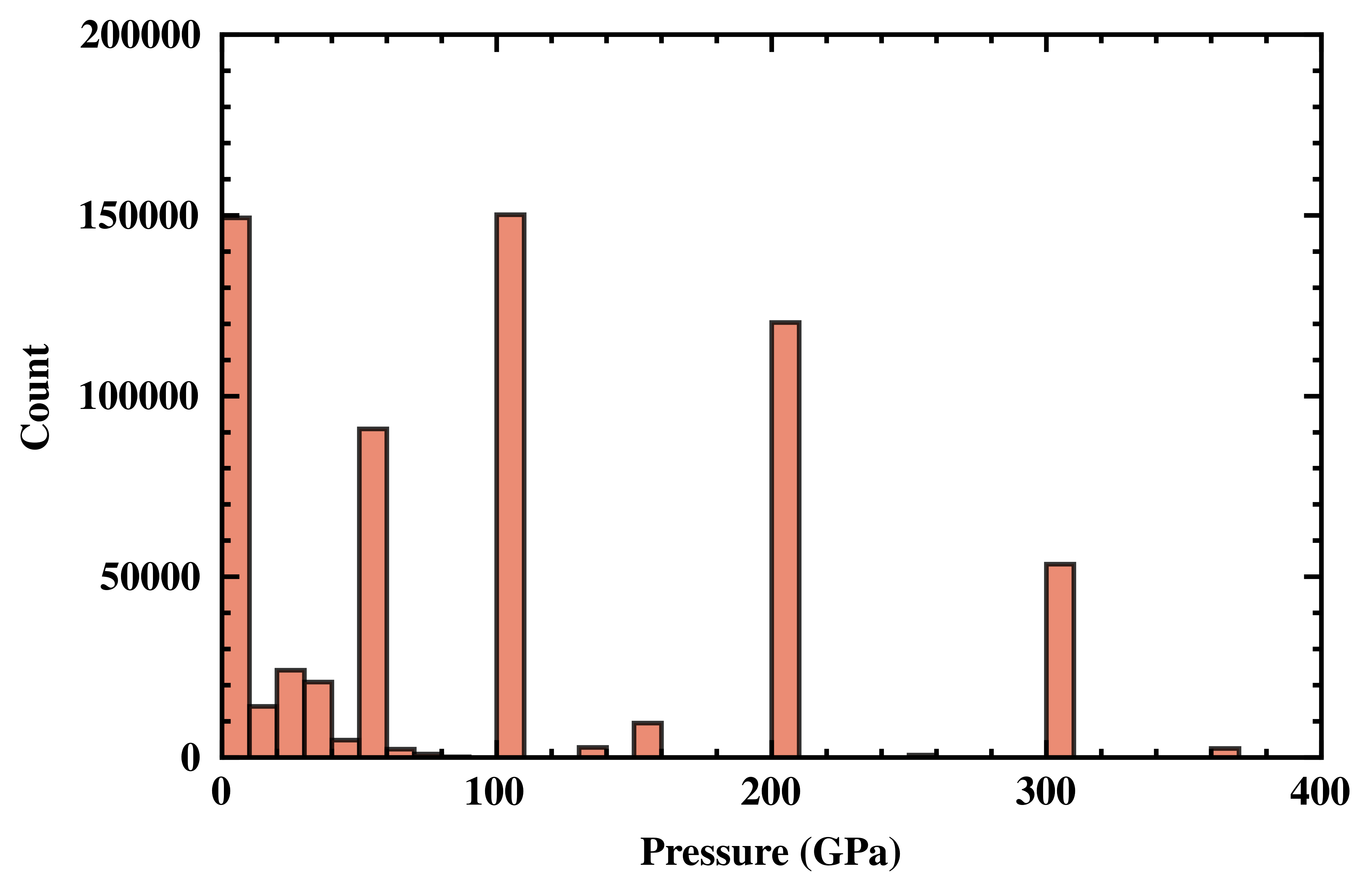}
\caption{\label{fig:s3-1} Distribution of equilibrium pressure in the MP-CALYPSO-60 dataset.}
\end{figure}

\clearpage

\section{\label{sec:s4}DFT computational details}
DFT calculations were performed with the Perdew-Burke-Ernzerhof (PBE) exchange-correlation functional~\cite{Perdew_PBE_1996} and all-electron projector-augmented wave method~\cite{Blochl_PAW_1994}, as implemented in the VASP code~\cite{Kresse_VASP_1996}. An energy cutoff of 520 eV and a Monkhorst-Pack k-point sampling grid spacing of 0.25 \AA\(^{-1}\) were used to ensure the convergence of the total energy. The default settings of PBE functional, Hubbard U corrections, and ferromagnetic initializations in Pymatgen~\cite{Ong_pymatgen_2013} \texttt{\detokenize{MPRelaxSet}} function were employed, except opting for the \texttt{\detokenize{W_sv}} potential for tungsten. The maximum optimization ionic step and the maximum running time are constrained to 150 steps and 20 hours, respectively.

\clearpage

\section{\label{sec:s5}Coverage metric for de novo generation}

Following previous works, we employ the coverage metric to measure the structural and compositional similarity between the test set \(\mathcal{S}_t\) and the generated structures \(\mathcal{S}_g\). Specifically, let \(d_S(\mathcal{M}_1, \mathcal{M}_2)\) and \(d_C(\mathcal{M}_1, \mathcal{M}_2)\) denote the L2 distances of the CrystalNN~\cite{Zimmermann_CrystalNN_2020} structural fingerprints and the normalized Magpie~\cite{Ward_Magpie_2016} compositional fingerprints. The Coverage Recall is defined as:

\[
\text{Cov-R} = \frac{1}{|\mathcal{S}_t|} \Big| \big\{ \mathcal{M}_i \,|\, \mathcal{M}_i \in \mathcal{S}_t, \exists \mathcal{M}_j \in \mathcal{S}_g, d_S(\mathcal{M}_i, \mathcal{M}_j) < \delta_S, d_C(\mathcal{M}_i, \mathcal{M}_j) < \delta_C \big\} \Big|,
\]

\noindent where \(\delta_S\) and \(\delta_C\) are pre-defined thresholds. The Coverage Precision is defined similarly by swapping \(\mathcal{S}_t\) and \(\mathcal{S}_g\). The recall metrics measure how many ground-truth materials are correctly predicted, while the precision metrics measure how many generated materials are of high quality.

\clearpage
\section{\label{sec:s6}Hyperparameters table for CrystalFlow}

\begin{table}[!ht]
\caption{\label{tab:s6-1}{Hyperparameters of CrystalFlow model used in this work} }
\begin{threeparttable}
\begin{tabular}{
l
S[output-exponent-marker=e,exponent-product=,table-align-exponent]
S[output-exponent-marker=e,exponent-product=,table-align-exponent]
}
\toprule
    {} &
    {\fontseries{b}\selectfont \quad CrystalFlow (CSP)  \quad} &
    {\fontseries{b}\selectfont \quad CrystalFlow (DNG) \quad} \\
\midrule
    {\fontseries{b}\selectfont Model} \\
\cmidrule(r){1-1}
Element type encoding dimension                             & 128               & 28 \\  
Gaussian expansion for pressure, number of bases            & 80                & 80  \\  
Gaussian expansion for pressure, bases start                & -2.0              & -2.0 \\  
Gaussian expansion for pressure, bases stop                 & 5.0               & 5.0 \\  
Time sinusoidal positional encoding dimension               & 256               & 256 \\
GNN hidden dimension                                        & 512               & 512   \\  
GNN number of layers ({\(L\)})                              & 6                 & 6   \\  
Number of frequency for {\(\mathbf{F}_{ij}^{\text{FT}}\)}         & 256               & 256 \\  
Loss weight \(\lambda_{k}\)                              &  1                  &  1 \\  
Loss weight \(\lambda_{F}\)                              & 10                  & 10 \\  
Loss weight \(\lambda_{A}\)                              & {--}                & 10 \\  
\toprule
    {\fontseries{b}\selectfont Optimizer} & &\\
\cmidrule(r){1-1}
Optimizer type                          & {Adam}                & {Adam} \\
Learning rate                           &  1e-3                 &  1e-3 \\
Learning rate scheduler                 & {ReduceLROnPlateau}   & {ReduceLROnPlateau} \\
Scheduler patience (epoch)              & 30                    & 30 \\
Scheduler factor (epoch)                & 0.6                   & 0.6 \\
Minimal learning rate                   & 1e-5                  & 1e-5 \\
\bottomrule
\end{tabular}
\end{threeparttable}
\end{table}

Each element type is encoded by a embedding layer into a vector of length 128 in CSP task, or a concatnated one-hot vector of length 28 in DNG task. Pressures are unit in GPa, and are first standardized on training set, then expanded to 80 equidistant Gaussian bases centered from -2.0 to 5.0 with standard deviation being the interval, i.e., \(7/80\)\,. The training batch size is 256 for MP-20 dataset, and 128 for both MPTS-52 and MP-CALYPSO-60 datasets.

\clearpage
\section{\label{sec:s7}Inference time compared to other generative model}

\begin{table}[!ht]
\caption{\label{tab:s7-1}The time is evaluated on MP-20 testset and reduced to the generation of 10,000 materials. Best results are marked in bold. Records of FlowMM and FlowLLM are from Ref.~\cite{Sriram_FlowLLM_2024}}.
\begin{threeparttable}
\begin{tabular}{
ll
S[table-format=4.0,detect-weight,mode=text]
S[table-format=2.1,detect-weight,mode=text]
}
\toprule
            & {Device} & \text{\(S\)} & \text{Time (min/10k)} \\
\midrule
DiffCSP     & RTX 4090  & 1000 & 44.7 \\
FlowMM      & A800      &  750 & 65.1 \\
FlowLLM     & A800      &  250 & 89.6 \\
CrystalFlow & RTX 4090  &  100 & \fontseries{b}\selectfont  4.1 \\
CrystalFlow & RTX 4090  & 1000 & \fontseries{b}\selectfont 37.0 \\
\bottomrule
\end{tabular} 
\end{threeparttable}
\end{table}

\end{widetext}

\end{document}